\documentstyle[12pt,epsf]{article}

\newcommand{\be}{\begin{equation}}
\newcommand{\ee}{\end{equation}}
\newcommand{\bea}{\begin{eqnarray}}
\newcommand{\eea}{\end{eqnarray}}
\newcommand{\ba}{\begin{array}}
\newcommand{\ea}{\end{array}}
\newcommand{\journal}[4]{{\rm #1} {\bf #2} (19#3) #4}
\newcommand{\NP}{\journal{Nucl. Phys.}}
\newcommand{\PL}{\journal{Phys. Lett.}}
\newcommand{\PRL}{\journal{Phys. Rev. Lett.}}
\newcommand{\PR}{\journal{Phys. Rev.}}
\newcommand{\NPPS}{\journal{Nucl. Phys. Proc. Suppl.}}

\newcommand{\JHEP}{\journal{JHEP}}
\newcommand{\MPL}{\journal{Mod. Phys. Lett.}}
\newcommand{\PTP}{\journal{Prog. Theor. Phys.}}

\begin{document}

\begin{titlepage}
\begin{flushright}
       {\normalsize  OU-HET 314 \\  hep-th/9903081\\
           March. 1999}
\end{flushright}
%
\begin{center}
  {\large \bf String creation in D6-brane background}

\renewcommand{\thefootnote}{\fnsymbol{footnote}}
\vfill
         {\bf Toshihiro ~Matsuo}\footnote{e-mail address: matsuo@funpth.phys.sci.osaka-u.ac.jp}  \\
            and \\
         {\bf Takashi ~Yokono}\footnote{e-mail address: yokono@funpth.phys.sci.osaka-u.ac.jp}\\
\vfill
        Department of Physics,\\
        Graduate School of Science, Osaka University,\\
        Toyonaka, Osaka, 560 Japan\\
\renewcommand{\thefootnote}{\arabic{footnote}}
\end{center}
\vfill
\begin{abstract}
The production of string charge during a crossing of certain oriented D-branes
is studied.
We compute the string charge in the system of a probe D2-brane and a background D6-brane by use of the equations of motion
in the ten-dimensions.
We confirm the creation of string charge as inflow from the
background D6-brane.
\end{abstract}
\vfill

\end{titlepage}

It was first pointed out in \cite{HW} that D3-branes are created
when certain oriented NS5-brane and D5-brane cross each other
and then the created branes are suspended between these 5-branes.
This phenomenon has several cousins related by U-duality.
For example, a fundamental string is created when a
Dp-brane and a D(8-p)-brane, which are mutually transverse, cross.
They have been confirmed from various points of view
\cite{BDG}-\cite{CGS}.

However, it has not been clarified yet how the charge of created brane
emerges when the branes cross.
In this letter we examine in particular the system of a probe D2-brane
and a background D6-brane. We calculate the created string charge by use of
the equations of motion derived from the action in the ten-dimensions.
This ten-dimensional action itself is obtainable from the eleven-dimensional
one with M2-brane source term.
The D6-brane background is realized by the Taub-NUT geometry in the eleven-dimensions.
The string charge in this particular system
turns out to be $Q_{F1} = \pi RT(1+\cos\theta_0 )$,
where $\theta_{0}$ is a parameter of the M2-brane embedding,
$T$ is the tension of it and $R$ is the radius of $S^1$ of the eleventh-direction. 
Before or after the crossing of these branes (equivalently $\theta_0$ is $\pi$ 
or $0$ respectively, as will be explained later), one can see $Q_{F1}$ will be $0$ or $2\pi RT$.
This means indeed the creation of string charge which is supplied from the
background D6-brane.

The original technique used here was given in \cite{GHM} in which they studied the duality between KK- and H-monopoles in the type II theory by showing the
charge inflow of a string into the KK-monopole background.
\\

Our starting point is the eleven-dimensional SUGRA (M-theory) whose
bosonic part is \cite{D}
\begin{eqnarray}
S_{11D}
=
\frac{1}{2\kappa_{11}^2}\int d^{11}x \sqrt{-{\cal{G}}}
\left[
R_{11} - \frac{1}{2 \cdot 4!}G_{MNPQ}^{2}
\right]
-\frac{1}{2\kappa_{11}^2 \cdot 6}\int C_3 \wedge G_4 \wedge G_4 ,
\label{a1}
\end{eqnarray}
where ${\cal G}_{MN}$ is the eleven-dimensional metric
($M, N, \ldots = 0, 1, \ldots, 10$)
and $G_4=dC_3$ is the field strength of the three-form gauge field
$C_3$.
The M2-brane has the electric charge of $C_3$.
Adding the non-linear $\sigma $-model action as the source term of
M2-brane to the above action, it becomes 
$S_{11D}+S_{11D}^{M2source}$
where
\begin{eqnarray}
S_{11D}^{M2source} = T \int d^{11}x \int d^{3} \xi
\delta^{(11)}(x^{M} -  X^{M})
\{
-\frac{1}{2} \sqrt{\gamma}\gamma^{ij}\partial_{i}X^{M}
\partial_{j}X^{N}{\cal{G}}_{MN}
\nonumber \\
+ \frac{1}{2} \sqrt{\gamma}
+ \frac{1}{3!}\epsilon^{ijk}
\partial_{i}X^{M}\partial_{j}X^{N}\partial_{k}X^{P}C_{MNP}
\}. \label{a2}
\end{eqnarray}
T is the membrane tension and
$X^{M}$ are the space-time coordinates of the M2-brane.
The world volume coordinates and the induced metric on the M2-brane
are denoted by $\xi^i ( i=0,1,2 )$ and $\gamma_{ij}$.

Let us consider the dimensional reduction of the actions into the
ten-dimensions. They provide the IIA SUGRA theory.
The eleven-dimensional metric ${\cal G}_{MN}$ transmutes into
ten-dimensional metric $g_{\mu\nu}$ ($\mu, \nu = 0, 1, \ldots , 9$),
one-form gauge potential $A$ and dilaton field $\phi$:
\be
{\cal G}_{MN} = e^{-\frac{2}{3}\phi}\left( \ba{cc}
g_{\mu\nu} + e^{2\phi}A_\mu A_\nu & e^{2\phi} A_\mu \\
e^{2\phi} A_\nu & e^{2\phi}
\ea \right). \label{mt}
\ee
$C_3$ provides three-form $C_3^{10}$ and two-form $B_2^{10}$ by the
reduction;
\bea
C_{\mu\nu\rho} &=& C^{10}_{\mu\nu\rho},  \label{4} \\
C_{\mu\nu 10} &=& B^{10}_{\mu\nu}.
 \label{5}
\eea
We will abbreviate $C_3^{10}$ and $B_2^{10}$ to $C_3$ and $B_2$.
Inserting (\ref{mt}), (\ref{4}) and (\ref{5}) into the actions, we
obtain
$S_{10D}^{bulk}+S_{10D}^{source}$ where
\begin{eqnarray}
S_{10D}^{bulk} = \int d^{10}x\sqrt{-g}
\left[
e^{-2\phi}
\left\{ R_{10}+4(\partial \phi)^{2}
-\frac{1}{2\cdot 3!}H_{\mu\nu\rho}^{2}\right\} \right.
\nonumber \\
\left.
- \frac{1}{2\cdot 2!}F_{\mu\nu}^{2}
- \frac{1}{2\cdot 4!}G_{\mu\nu\rho\sigma}^{2}
\right]
- \frac{1}{2} \int dC_3\wedge dC_3 \wedge B_2,
\label{10dbulk}
\end{eqnarray}
and
\begin{eqnarray}
S_{10D}^{source} =  T \int d^{10}x \int d^{3} \xi
\delta^{(10)}(x^{\mu} - X^{\mu})
\{
 \frac{1}{2} \sqrt{\gamma}
-\frac{1}{2} \sqrt{\gamma}\gamma^{ij}e^{-\frac{2}{3}\phi}
 (h_{ij} - e^{2 \phi }V_{i}V_{j})
 \nonumber \\
 + \frac{1}{3!}\epsilon^{ijk}
 \partial_{i} X^{\mu} \partial_{j} X^{\nu} \partial_{k} X^{\rho}
C_{\mu\nu\rho}
 + \frac{1}{2}\epsilon^{ijk}
 \partial_{i}X^{\mu}\partial_{j}X^{\nu}(V_{k} - \partial_k X^{\rho}
A_{\rho})B_{\mu \nu}
 \}.
 \label{10dsource}
\end{eqnarray}
In (\ref{10dbulk}), $F_2 = d A_1, H_3 = d B_2 $
and $G_4 = d C_3 + A_1 \wedge H_3 $.
We take $2\kappa_{10}^2 =1$.
The source term of M2-brane reduces to (\ref{10dsource}), where $V_i$ and
$h_{ij}$ are
\begin{equation}
V_{i} = \partial_{i}X^{10} + \partial_i X^{\mu} A_{\mu},\quad
h_{ij}=\partial_{i}X^{\mu}\partial_{j}
X^{\nu}g_{\mu\nu}.
\end{equation}
The equations of motion which we need are those for the
fields $C,B$ and $A$.
They can be read from (\ref{10dbulk}) and (\ref{10dsource})
as follows;
\begin{eqnarray}
&C&: \qquad
\partial_{\sigma}
(\sqrt{-g}G^{\sigma \mu \nu \rho}
+\frac{1}{2!\cdot 3!}\epsilon^{\sigma \mu \nu \rho \dots \dots}
\partial_{.}C_{...}B_{..}
)
\nonumber \\
& &  \qquad \qquad \qquad \qquad
=
-T \int d^{3} \xi
\delta^{(10)}(x^{\mu} - X^{\mu})
\epsilon^{ijk}\partial_{i}X^{\mu}\partial_{j}
X^{\nu}\partial_{k}X^{\rho},
\nonumber \\
&B&: \qquad
\partial_{\rho}
(\sqrt{-g}e^{-2 \phi}H^{\rho \mu \nu}
+  \sqrt{-g} G^{\rho \mu \nu \sigma} A_{\sigma}
) + \frac{1}{2\cdot 3!\cdot 3!} \epsilon^{\mu \nu \cdots \cdots
\cdot \cdot}
\partial_{.}C_{...}\partial_{.}C_{...}
\nonumber \\
& &  \qquad \qquad
=
-\int d^{3} \xi \delta^{(10)}(x^{\mu} - X^{\mu})
\epsilon^{ijk}\partial_{i}X^{\mu}\partial_{j}X^{\nu}(V_{k}-A_{k}),
\nonumber \\
&A&: \qquad \partial_{\nu}(\sqrt{-g}F^{\nu \mu}) - \frac{1}{6}
\sqrt{-g} G^{\mu\nu\rho\sigma}H_{\nu\rho\sigma} \label{eqmotion0} \\
&& \qquad \qquad =
T \int d^{3} \xi \delta^{(10)}(x^{\mu} - X^{\mu})
\sqrt{\gamma}\gamma^{ij}e^{\frac{4}{3}\phi}\partial_{i}X^{\mu}V_{j}.
\nonumber
\end{eqnarray}

Our next task is to describe a D6-brane background in the IIA
theory.
It is well known that the action (\ref{a1}) admits the
eleven-dimensional KK-monopole solution and in the IIA limit
it becomes the D6-brane \cite{T}.
Thus we consider the KK-monopole in the eleven-dimensions
which is described by Taub-NUT metric (plus the flat
seven-dimensions),
\begin{eqnarray}
ds^{2}
=\eta_{mn}dx^{m}dx^{n} + U \left[ dx^{10} + A_\phi d\phi
\right]^{2}
+ U^{-1}(dr^{2} + r^{2}d\Omega^{2}_{2}),
\label{TNmetric}
\end{eqnarray}
where $m,n=0, 1, \ldots, 6$.
To avoid the singularity at $r=0$, $x^{10}$ must have periodicity $2
\pi R$,
where R is the radius of the circle of the eleventh-direction.
The potential terms $A_\phi$ and $U(r)$ are
\bea
A^{S}_\phi &=&\frac{R}{2}(1 + \cos\theta), \label{south}\\
U(r) &=& (1 + \frac{R}{2r})^{-1}.
\eea
We denote the coordinates regular at the south pole ($\theta = \pi$) 
and the north pole ($\theta =0$) respectively by the indices $S$ and $N$.
The metric (\ref{TNmetric}) defined by eq.(\ref{south}) is
non-singular at  the south pole but has a coordinate
singularity at the north pole($\theta = 0$).
The singularity can be removed by
\bea
x_{N}^{10} &=& x_{S}^{10} + R\phi , \label{gauge}
\eea
then the direction of Dirac string is changed,
\bea
A_\phi^{N} &=& \frac{R}{2}(-1+\cos \theta ).
\eea
This transformation of coordinates plays particularly important role
when we discuss the creation of string charge.
Comparing the metric to eq.(\ref{mt}), we take $ U=e^{\frac{4}{3}\phi}$.

The equations (\ref{eqmotion0}) could become simpler in the Taub-NUT background.
The term in the left hand side of the third equation in
(\ref{eqmotion0}), $\partial_{\mu}\left(
\sqrt{-g}F^{\mu\nu}\right)$,
trivially vanishes.
The terms in the eqs.(\ref{eqmotion0}), which originate from
Chern-Simon-like term in the action (\ref{10dbulk}),
turn out to be irrelevant as will be shown.
Dropping out these terms, the equations of motion (\ref{eqmotion0})  
become
\bea
C&:&\partial_{\sigma} (\sqrt{-g}G^{\sigma\mu\nu\rho})
=\sqrt{-g}j_{D2}^{\mu\nu\rho }
\equiv - T\int d^3 \xi \delta^{(10)}(x^\mu
-X^\mu)\epsilon^{ijk}\partial_i X^\mu \partial_j X^\nu \partial_k
X^\rho , \nonumber\\
B&:&\partial_{\rho }h^{\rho\mu\nu }
=\sqrt{-g} j_{F1}^{\mu\nu} \label{eqmotion} \nonumber \\
& &\quad \equiv   - \int d^{3} \xi \delta^{(10)}(x^{\mu} - X^{\mu})
\epsilon^{ijk}\partial_{i}X^{\mu}\partial_{j}X^{\nu}\partial_k
X^{10}
+\partial_\rho (\sqrt{-g}G^{\rho\mu\nu\phi}A_\phi ),
\nonumber \\
A&:&G^{\mu\nu\rho\sigma}h_{\nu\rho\sigma} = 6T \int d^{3} \xi
\delta^{(10)}(x^{\mu} - X^{\mu})
\sqrt{\gamma}\gamma^{ij}U^{-\frac{1}{2}}\partial_{i}X^{\mu}V_{j} ,
\eea
where $h^{\rho\mu\nu}\equiv
\sqrt{-g}U^{-\frac{3}{2}}H^{\rho\mu\nu}$, $j_{D2}$ and $j_{F1}$ are the D2-brane and the fundamental string currents. We take p-brane charge as  $Q=\int_{S^{8-p}}*F_{p+2}=\int_{V^{9-p}}*j_{p+1}$.
Here $F_{p+2}$ is the $(p+2)$-form field strength and $j_{p+1}$ is
the p-brane current. The definition of the Hodge dual contain
a dilaton dependent factor to correctly have a generalized
electric-magnetic duality.
\pagebreak[2]

Now, let us consider the following embedding of a probe M2-brane
in the eleven-dimensions in order to check the consistency of
eqs.(\ref{eqmotion}),
\begin{eqnarray}
X^{10} &=& 2 \pi R \xi^{1},
\nonumber \\
X^{t}  &=& \xi^{0},
\nonumber \\
X^{r}  &=&  \xi^{2},
\nonumber \\
X^{\theta}  &=& 0,
\nonumber \\
X^{\phi}  &=& 0,
\nonumber \\
X^{m}  &=& 0 \qquad m=1\ldots 6.
\end{eqnarray}
This embedding describes the configuration of an M2-brane
wrapping exclusively around the eleventh-direction.
Therefore, in the ten-dimensions, it becomes a string
which is extending
towards the $r$-direction.

With this embedding the eqs.(\ref{eqmotion}) become
\begin{eqnarray}
&C&: \qquad \partial_{\sigma}(\sqrt{-g}G^{\sigma \mu \nu \rho})=0,
\nonumber \\
&B&: \qquad
\partial_{\rho} h^{\rho \mu \nu} = \sqrt{-g} j_{F1}^{\mu\nu}
\nonumber \\
& &  \qquad \qquad \quad
= \partial_\rho (\sqrt{-g}G^{\rho\mu\nu\phi}A_\phi ) - 2\pi RT \int
d^{3} \xi \delta^{(10)}(x^{\mu} - X^{\mu})
\epsilon^{ij1}\partial_{i}X^{\mu}\partial_{j}X^{\nu},
\nonumber \\
&A&:  \qquad \qquad
G^{\mu \nu \rho \sigma}h_{\nu \rho \sigma}
= 0.
\end{eqnarray}
Let $G^{\mu\nu\rho\sigma}=0$.
The current which couples to the NS-NS
$2$-form $B$ can be read as
\begin{equation}
\sqrt{-g}j^{tr}_{F1}
=
RT \delta(\theta)\delta(\phi)\delta^{(6)}(x).
\end{equation}
Thus We have the charge of the fundamental string,
\begin{eqnarray}
Q_{F1}
&=&
\int d^{6}x d\theta d\phi \sqrt{-g}j^{tr}
\nonumber \\
&=& 2\pi R T.
\end{eqnarray}
This is in agreement with the relation of string charge
and M2-brane charge. Hence the equations (\ref{eqmotion}) provide the desired
result for the above embedding.\\

We would like to consider the system of a probe D2-brane in the D6-brane
background both of which are mutually transversed.
We will show the creation of string charge
in view of the IIA theoretical stand point.
For this purpose we examine the following embedding of an M2-brane in the
eleven-dimensions.
\begin{eqnarray}
X^{t}  &=& \xi^{0},
\nonumber \\
X^{r}  &=& \xi^{2},
\nonumber \\
X^{\theta}  &=& \theta_{0} \mbox{(const)},
\nonumber \\
X^{\phi}  &=& 2 \pi \xi^{1},
\nonumber \\
X^{m}  &=& 0,
\nonumber \\
X_{S}^{10} &=& 0.
\label{embedding}
\end{eqnarray}
Obviously, the M2-brane has no winding around the eleventh-direction in the coordinate system regular at the south pole.
On the other hand, it can wind around the direction,
i.e. $X_N^{10}=2\pi\xi^1 $, in the coordinate system regular at
the north pole obtained by the gauge transformation
(\ref{gauge}). This seems to imply string charge creation
in the ten-dimensional view point.
We will investigate concretely this phenomenon by use of the
equations of motions(\ref{eqmotion}).

\begin{figure}[t]
\epsfysize=6cm \centerline{\epsfbox{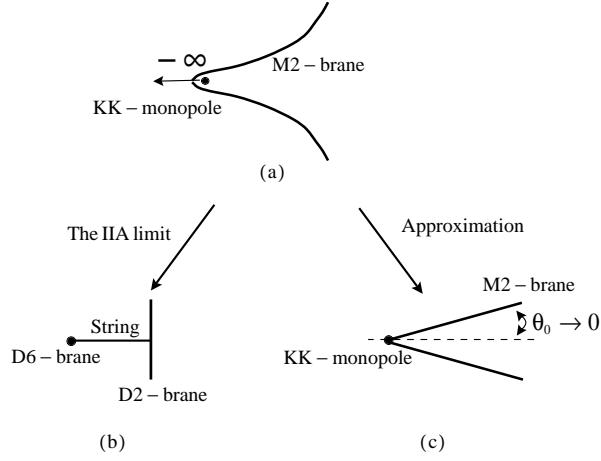}}
\caption{\small
(a)The M2-brane holomorphically embedded in a KK-monopole
background.
(b)The string which is created in the IIA limit.
(c)The surface approximated by a cone.
}
\label{fig}
\end{figure}

Our interest is the string charge that is the topological charge
associated with the winding around the eleventh-direction.
Hence we do not need to take care of supersymmetry.
However, before calculating the string charge, it is worth considering the
correspondence between the embedding (\ref{embedding}) and that of \cite{NOYY,Y}
which preserve supersymmetry manifestly. In those papers
string creation was shown by investigating geometrically the shape of
an M2-brane but not the winding around the eleventh-direction.
They introduce \cite{NOYY2} the following holomorhpic coordinates $(v,y)$ to describe the embedding of the M2-brane.
\bea
y&=&e^{-\frac{1}{R}(r\cos\theta +ix_s^{10})}\sqrt{r(1-\cos\theta)} ,  \\
v&=&\frac{2}{R}r\sin\theta e^{i\phi}.
\eea
M2-brane embedded holomorphically in the KK-monopole background is
given by
\footnote{
Although the configuration considered in \cite{NOYY} is M5-brane
with a KK-monopole, the same relation is also applicable to our
case, an M2-brane, because the condition of holomorphy is for the
two-dimensional part of branes. In our case  the topology of the
M2-brane world volume is $R^1\times \Sigma $, where $\Sigma $ is
holomorphically embedded Riemann surface in Taub-NUT space.
}
\be
y = const. \label{yv}
\ee
The shape of the M2-brane described by this curve looks a
parabolic-like surface, see Fig \ref{fig}(a).
The embedding (\ref{embedding}) means
that we approximate this surface by a cone in the eleven-dimensions, see Fig \ref{fig}(c).
Even though this approximation is not compatible with complex structure of the Taub-NUT space, we may use the approximation to study the
NS-NS charge since the embedding to the eleventh-direction will play the role
in the determination of the charge and it can be still captured by this approximation.
The KK-monopole is located in the distance of the right (left) infinity
in the eleven-dimensions before (after) string creation in ten-dimensions
\cite{NOYY,Y}.
Here we set the position of the KK-monopole the origin by a parallel shift
of the coordinates.
The configuration in Fig \ref{fig}(a), which describes the created string
in the ten-dimensions, corresponds to the case of $\theta_0 =0$.
The M2-brane should be pulled by the KK-monopole when it moves,
since it is attached to the M2-brane in the embedding (\ref{embedding}).
Hence when the KK-monopole moves to the right
direction in Fig \ref{fig}, $\theta_0$ should become $\pi$.
Therefore the parameter $\theta_0$ should be $\pi$ or $0$
before or after the branes cross, respectively.

We now start to calculate string charge by
inserting the embedding (\ref{embedding}) (or (\ref{gauge})) to
(\ref{eqmotion}).
As the probe brane is a point-like object in view of the space of
the D6 brane world volume, it is natural to take spherical coordinates,
$l$ and $\varphi_i \,\,(i=1,\ldots ,5 )$ where $l$
is the radial direction and $\varphi_i$'s are the angular
ones in the space. We assume that the indices run over only the
$t, r, \theta, \phi, l $ directions.
As we mentioned before, the parts in eqs.(\ref{eqmotion0}) from the
Chern-Simon term automatically vanish under the assumption.
We find the following solution,
\bea
\sqrt{-g} G^{t r \phi l}
&=& - T \delta(\theta - \theta_{0})
\Theta(l)
\delta^{(5)}(\varphi^i), \nonumber \\
h^{tr\theta}
&=&
2RT(1+\cos{\theta_{0}})\Theta(\theta - \theta_{0})
\delta (l)\delta^{(5)}(\varphi^i)
- \alpha(r,\theta_{0})\delta^{'}(x^{l})\cdot \theta ,
\nonumber \\
h^{trl}
&=&
\alpha(r,\theta_{0})\delta(x^{l}),
\nonumber \\
h^{t \theta l}
&=&
- \partial_{r}\alpha(r,\theta_{0})\delta(x^{l}) \cdot \theta ,
\nonumber \\
\mbox{otherwise} &=& 0 ,
\eea
where $\alpha(r,\theta) = -4\pi R r^{2}
\sin\theta(1+\cos\theta)\sqrt{|\gamma |} \gamma^{11}$ and we used
the fact that $\Theta(x)\delta(x) = \frac{1}{2} \delta(x)$.
Then the non-vanishing currents of the D2-brane and the string
become
\bea
\sqrt{-g}j_{D2}^{tr\phi} &=& T\delta (l)\delta^{(5)}(\varphi^i), \\
\sqrt{-g}j_{F1}^{tr} &=& \frac{RT}{2}(1+\cos\theta_0)\delta(\theta
-\theta_0)\delta (l) \delta^{(5)} (\varphi^i).
\eea
We should note that the form of these equations does not depend on
the selection of coordinate patches.
The charge of the D2-brane and the string become
\bea
Q_{D2} &=& T, \\
Q_{F1} &=& \pi RT(1+\cos\theta_0 ).
\eea
Thus we finally obtain string charge creation,
$Q_{F1}$ becomes $2\pi RT$ from $0$ as $\theta_0$ going to $0$ from $\pi$.
This gives correct relation between the string tension and
that of D2-brane.
The net charge must be conserved in the whole space, therefore we
conclude that the electric NS-NS charge must be supplied from the D6-brane.
Note that $Q_{D2}$ does not depend on
$\theta_0 $. This means the
D2-brane always exists on the process of string creation.

%
%

\section*{Acknowledgments}
The authors acknowledge Asato Tsuchiya, Toshio Nakatsu,
Kazutoshi Ohta and Yosuke Imamura for discussions.
They are grateful to Toshio Nakatsu for a
careful reading of the manuscript. They also thank the
organizers and participants of Niseko Winter School.


\end{document}